\documentstyle[12pt,epsfig]{article}
\begin{document}

\begin{center}

\Large{Tensor Self Energy in a Vector-Tensor
Model}
\end{center}

\begin{center}
\large{A. Buchel$^{1,2}$, F. A. Chishtie$^{1}$, M.
Gagn$\acute{e}$-Portelance$^{1}$, S. Homayouni$^{1}$ and \\ D. G.
McKeon$^{1}$}

\smallskip

{\small \it $^{1}$ Department of Applied Mathematics,\\ The
University of Western Ontario\\London, Ontario N6A 5B7, Canada
\\ $^{2}$ Perimeter Institute for Theoretical Physics Waterloo, Ontario
N2J 2W9, Canada  } {\small UWO-TH-06/21}
\end{center}

\vskip 1cm

\abstract{The tensor self energy is computed at one loop order in
a model in which a vector and  tensor interact in a way that
eliminates all tensor degrees of  freedom. Divergencies arise
which cannot be eliminated without introducing a kinetic term for
the tensor field which does not appear in the classical action. We
comment on a possible resolution of this puzzle.}

\vskip 1cm

\baselineskip 24pt

It has been shown [1] that if a vector field $W_{\mu}^a$ and an antisymmetric tensor field $\phi_{\mu\nu}^a$ have a Lagrangian
\begin{equation}
L=-\frac{1}{4}F_{\mu\nu}^aF_{\mu\nu}^a+\frac{1}{12}G_{\mu\nu\lambda}^a
G_{\mu\nu\lambda}^a +
\frac{m}{4}\epsilon_{\mu\nu\lambda\sigma}\phi_{\mu\nu}^a
F_{\lambda\sigma}^a+\frac{\mu^2}{8}\epsilon_{\mu\nu\lambda\sigma}\phi_{\mu\nu}^a
\phi_{\lambda\sigma}^a
\end{equation}
then classically the tensor field has no physical degrees of freedom and that only the transverse polarizations of the vector field are dynamical. (Here we have\begin{equation}
F_{\mu\nu}^a=\partial_{\mu}W_{\nu}^a-\partial_{\nu}W_{\mu}^a+g f^{abc}W_{\mu}^b
W_{\nu}^c \end{equation}
\begin{equation}
G_{\mu\nu\lambda}^a=D_{\mu}^{ab}\phi_{\nu\lambda}^b +D_{\nu}^{ab}\phi_{\lambda\mu}^b +D_{\lambda}^{ab}\phi_{\mu\nu}^b
 \end{equation}
 and
\begin{displaymath}
D_{\mu}^{ab}=\partial_{\mu}\delta^{ab}+g f^{apb}W_{\mu}^p).
\end{displaymath}
In ref.[2] the vector self energy is computed to one loop order in this model. There it is demonstrated that the sum
 of those diagrams which receive a contribution from the tensor field is zero.  This is consistent with the tensor having no physical degrees of freedom.\\
In this paper we consider the tensor self energy
$<{\phi^{a}}_{\alpha1\beta1}(p){\phi^{b}}_{\alpha2\beta2}(-p)>$ in
this same model using the Feynman rules provided in ref.[2].
The pole parts of each of the one loop diagrams is given in Figure (1), where\\
\begin{displaymath}
K = g^2f^{mna}f^{mnb}/\pi^2,
\end{displaymath}
\begin{displaymath}
A =
p_{\alpha_1}p_{\alpha_2}\delta_{\beta_1\beta_2}-p_{\alpha_1}p_{\beta_2}\delta_{\beta_1\alpha_2}+
p_{\beta_1}p_{\beta_2}\delta_{\alpha_1\alpha_2}-p_{\beta_1}p_{\alpha_2}\delta_{\alpha_1\beta_2}
\end{displaymath}
and
\begin{displaymath}
B_1 =
p_\lambda(p_{\alpha_1}\epsilon_{\lambda\beta_1\alpha_2\beta_2}-p_{\beta_1}\epsilon_{\lambda\alpha_1\alpha_2\beta_2})
\end{displaymath}\begin{displaymath}
B_2 =
p_\lambda(p_{\alpha_2}\epsilon_{\lambda\beta_2\alpha_1\beta_1}-p_{\beta_2}\epsilon_{\lambda\alpha_2\alpha_1\beta_1})
\end{displaymath}\begin{displaymath}
C =
\delta_{\alpha_1\alpha_2}\delta_{\beta_1\beta_2}-\delta_{\alpha_1\beta_2}\delta_{\beta_1\alpha_2}.
\end{displaymath}
\clearpage


 \begin{figure}
 \begin{centering}
\begin{tabular}{|c |c |c| } \hline
    DIAGRAM & SYMMETRY FACTOR & POLE AT $\epsilon=0$\\
    \hline
\raisebox{1cm}{A1} \epsfig{file=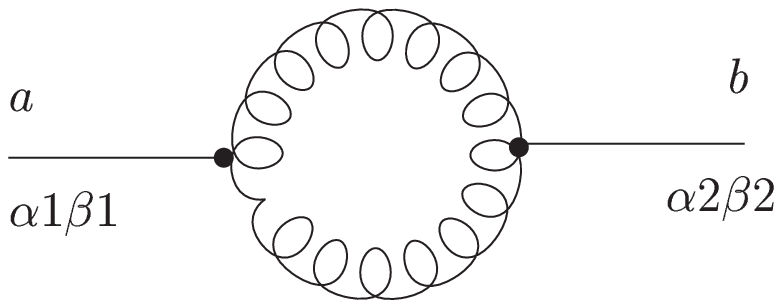, width =2in} &
\raisebox{1cm}{\parbox{.5in}{$1/2$}} &
\raisebox{.5in}{\parbox{2in}{$\frac{1}{16}\frac{m^2K}{p^2
\mu^8}[(p^2+m^2)^2 A-\\
\mu^2(p^2+m^2)(B_1+B_2)+p^2\mu^4C]$}}\\
\hline \raisebox{1cm}{A2} \epsfig{file=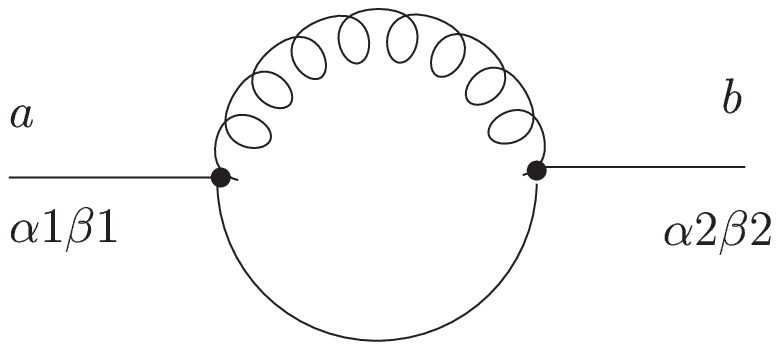, width =2in} &
\raisebox{1cm}{\parbox{.5in}{$1$}} &
\raisebox{.5in}{\parbox{2in}{$\frac{1}{96}\frac{p^2m^2K}{\mu^8}A$}}\\
\hline \raisebox{1cm}{A3} \epsfig{file=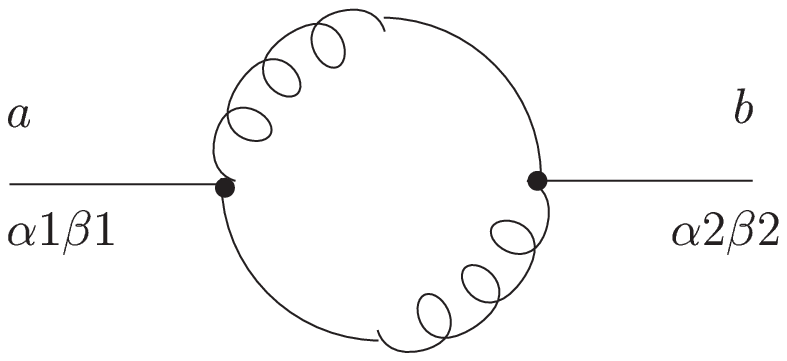, width =2in} &
\raisebox{1cm}{\parbox{.5in}{$1$}} &
\raisebox{.5in}{\parbox{2in}{$\frac{5}{96}\frac{p^2m^2K}{\mu^8}A$}}\\
\hline \raisebox{1cm}{A4a} \epsfig{file=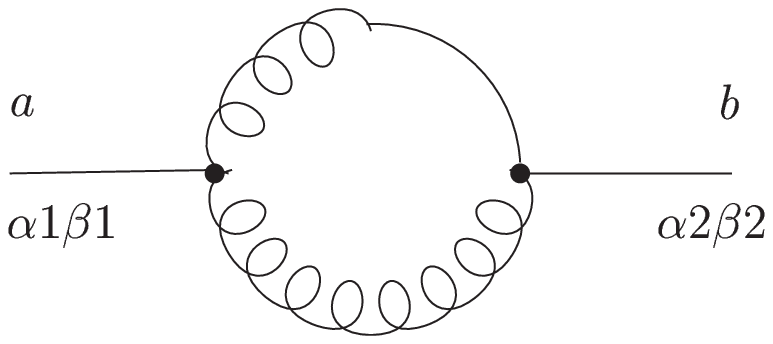, width =2in} &
\raisebox{1cm}{\parbox{.5in}{$1$}} &
\raisebox{.5in}{\parbox{2in}{$\frac{1}{16}\frac{m^2K}{\mu^8}((p^2+m^2)A-\mu^2B_2)$}}\\
\hline \raisebox{1cm}{A4b} \epsfig{file=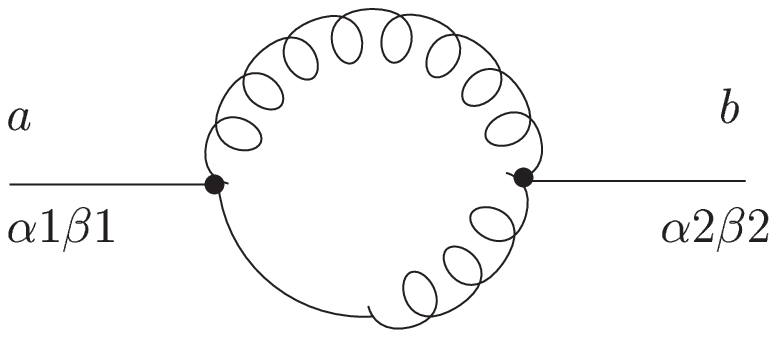, width =2in} &
\raisebox{1cm}{\parbox{.5in}{$1$}} &
\raisebox{.5in}{\parbox{2in}{$\frac{1}{16}\frac{m^2K}{\mu^8}((p^2+m^2)A-\mu^2B_1)$}}\\
\hline \raisebox{1cm}{B1} \epsfig{file=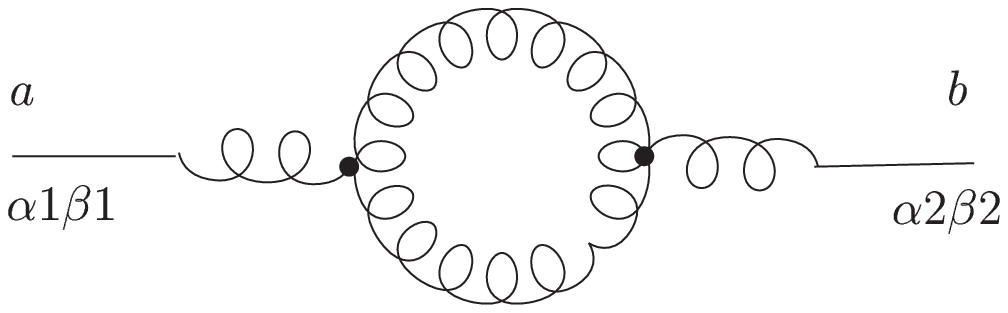, width=2in} &
\raisebox{1cm}{\parbox{.5in}{$1/2$}} &
\raisebox{.5in}{\parbox{2in}{$\frac{19}{192}\frac{m^2K}{p^2\mu^4}A$}}\\
\hline \raisebox{1cm}{B2} \epsfig{file=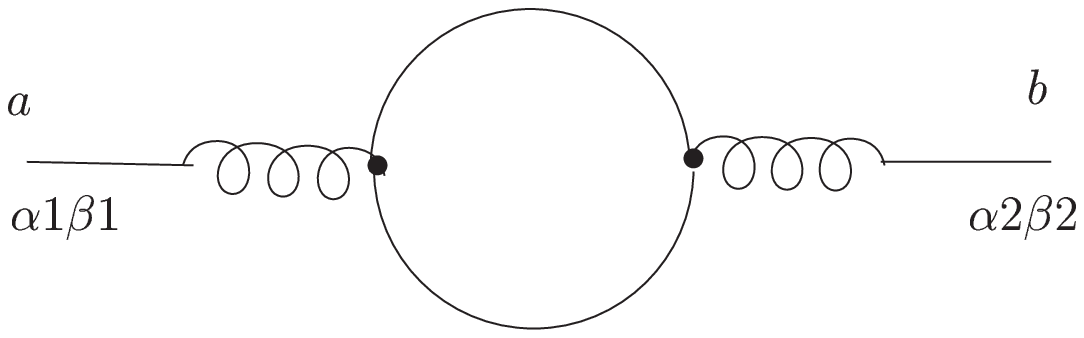, width =2in} &
\raisebox{1cm}{\parbox{.5in}{$1/2$}} &
\raisebox{.5in}{\parbox{2in}{$0$}}\\
\hline \raisebox{1cm}{B3} \epsfig{file=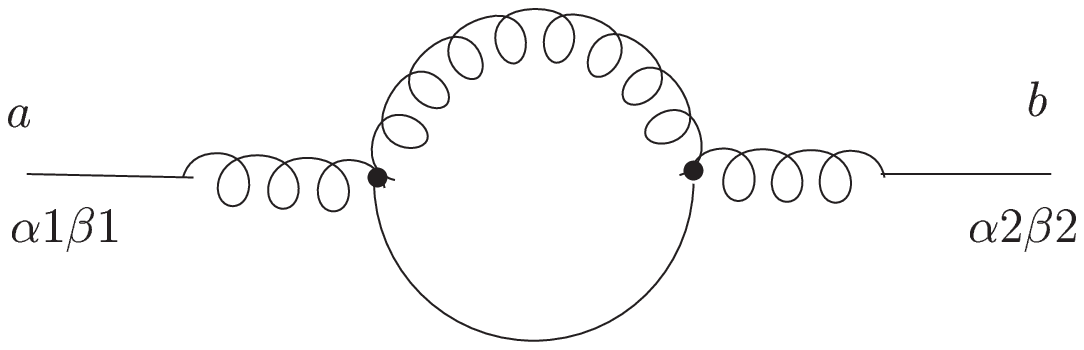, width =2in} &
\raisebox{1cm}{\parbox{.5in}{$1$}} &\raisebox{.5in}{\parbox{2in}{$\frac{1}{96}\frac{m^6K}{p^2\mu^8}A$}}\\
\hline
 \raisebox{1cm}{B4a} \epsfig{file=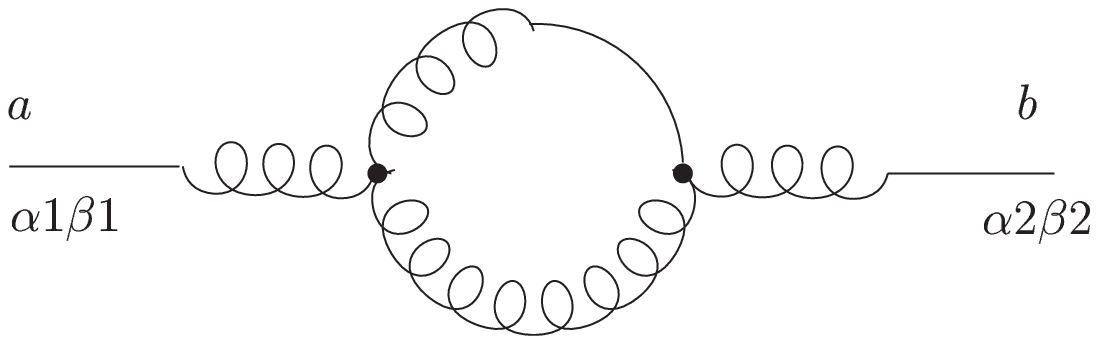, width =2in} &
\raisebox{1cm}{\parbox{.5in}{$1$}}
&\raisebox{.5in}{\parbox{2in}{$0$}}\\
 \hline \raisebox{1cm}{B4b}
\epsfig{file=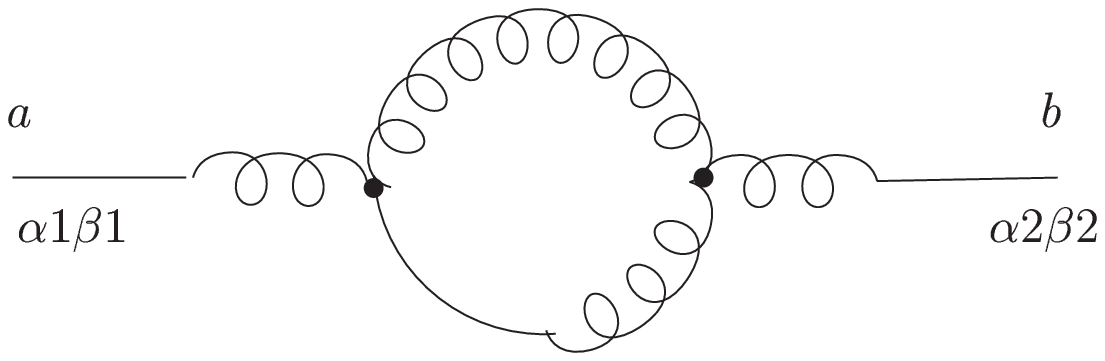, width =2in} &
\raisebox{1cm}{\parbox{.5in}{$1$}} &
\raisebox{.5in}{\parbox{2in}{$0$}}\\
\hline
\end{tabular}
\end{centering}
\end{figure}
\clearpage
\begin{figure}
 \begin{centering}
\begin{tabular}{|c |c |c| }
\hline
    DIAGRAM & SYMMETRY FACTOR & POLE AT $\epsilon=0$\\
    \hline
     \raisebox{1cm}{B5a} \epsfig{file=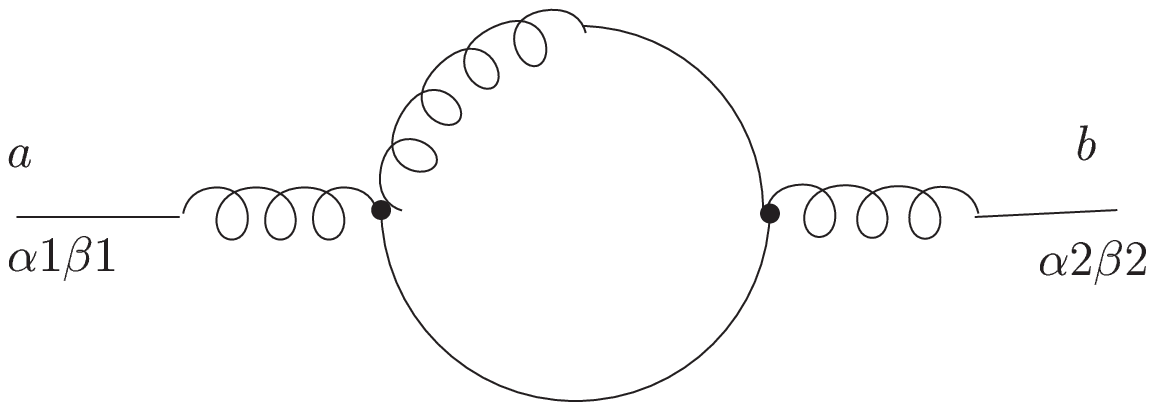, width =2in} &
\raisebox{1cm}{\parbox{.5in}{$1$}} &
\raisebox{.5in}{\parbox{2in}{$0$}}\\
\hline \raisebox{1cm}{B5b} \epsfig{file=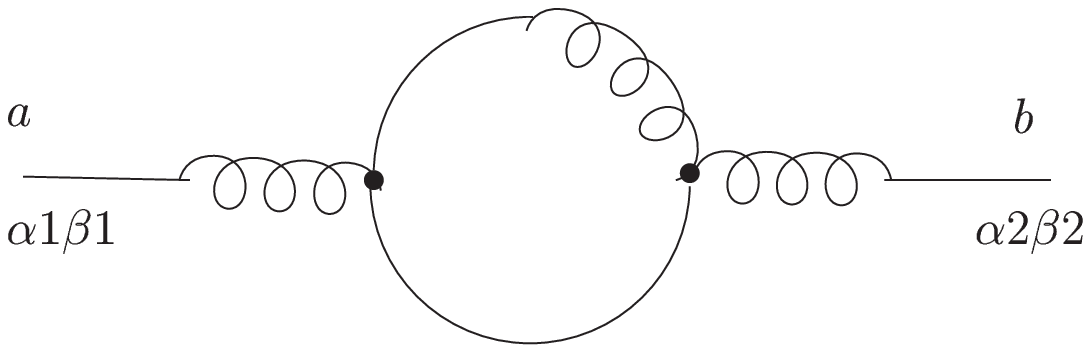, width =2in} &
\raisebox{1cm}{\parbox{.5in}{$1$}} &\raisebox{.5in}{\parbox{2in}{$0$}}\\

     \hline \raisebox{1cm}{B6} \epsfig{file=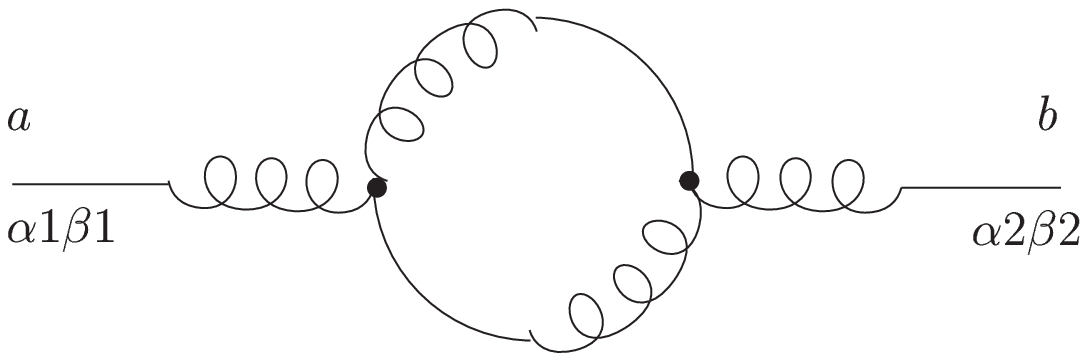, width =2in} &
     \raisebox{1cm}{\parbox{.5in}{$1$}} &\raisebox{.5in}{\parbox{2in}{$\frac {5}{96}\frac{m^6K}{p^2\mu^8}A$}}\\
     \hline \raisebox{1cm}{B7a} \epsfig{file=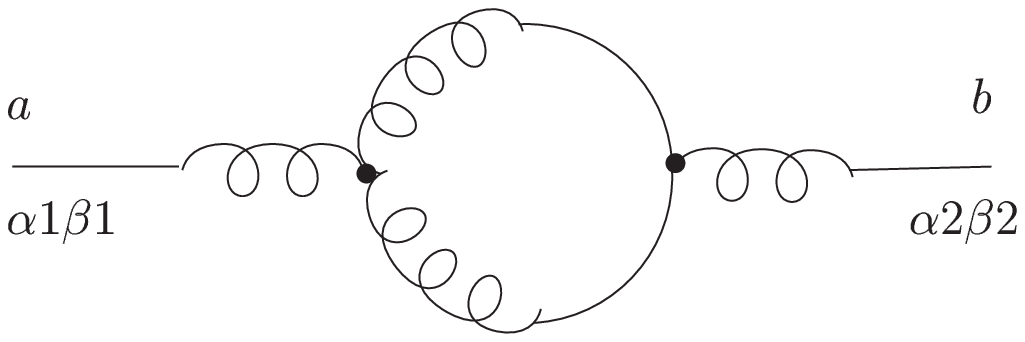, width =2in} &
\raisebox{1cm}{\parbox{.5in}{$1/2$}} &\raisebox{.5in}{\parbox{2in}{$0$}}\\
\hline \raisebox{1cm}{B7b} \epsfig{file=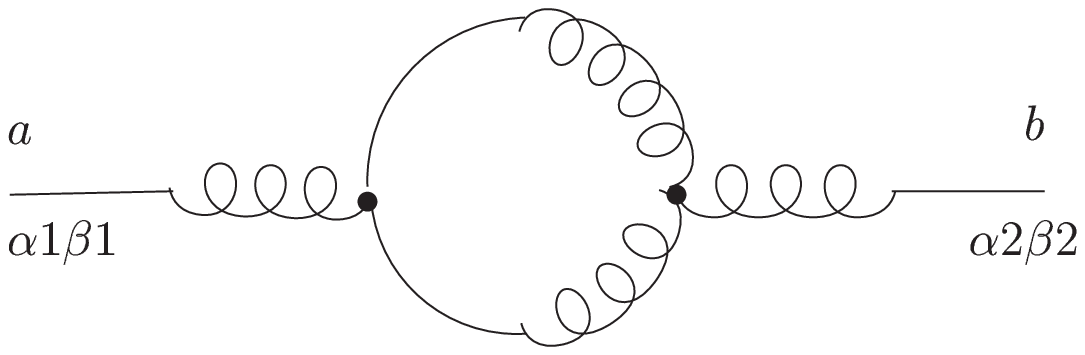, width =2in} &
\raisebox{1cm}{\parbox{.5in}{$1/2$}} &\raisebox{.5in}{\parbox{2in}{$0$}}\\
\hline \raisebox{1cm}{B8} \epsfig{file=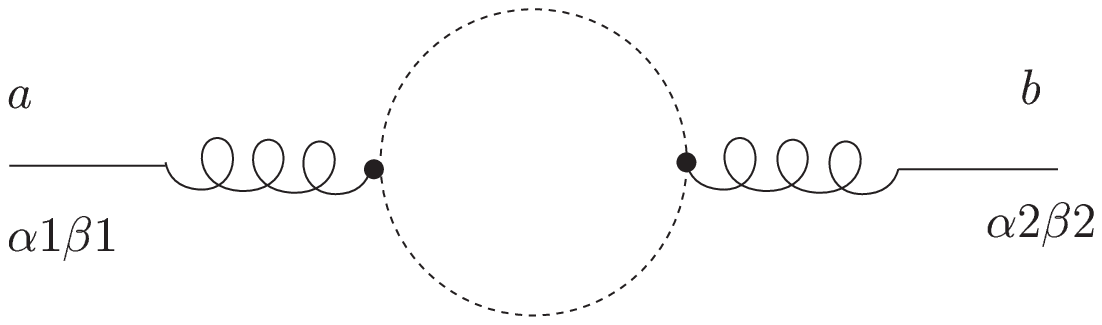, width =2in} &
\raisebox{1cm}{\parbox{.5in}{$-1$}}
&\raisebox{.5in}{\parbox{2in}{$\frac{1}{192}\frac{m^2K}{p^2\mu^4}A$}}\\
\hline \raisebox{1cm}{C1} \epsfig{file=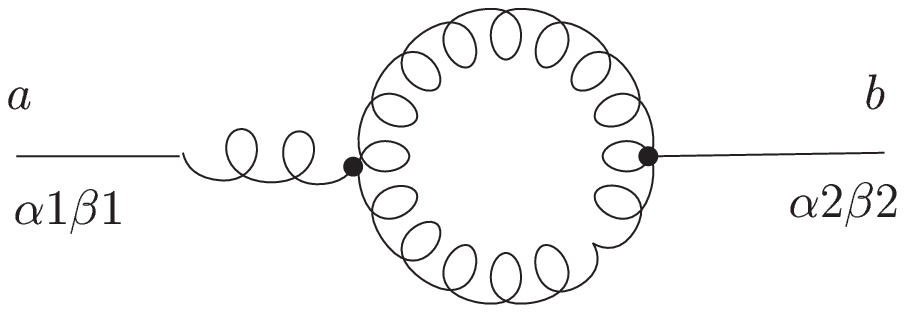, width =2in} &
\raisebox{1cm}{\parbox{.5in}{$1/2$}}
&\raisebox{.5in}{\parbox{2in}{$\frac{3}{32}\frac{m^2K}{p^2\mu^4}A$}}\\
\hline \raisebox{1cm}{C2} \epsfig{file=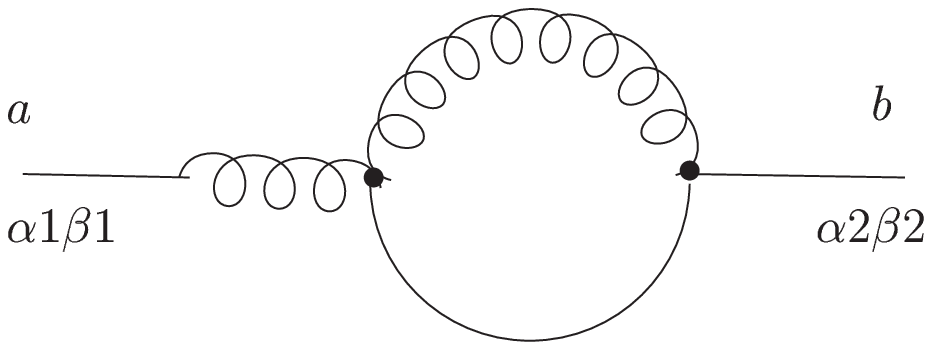, width =2in} &
\raisebox{1cm}{\parbox{.5in}{$1$}} &\raisebox{.5in}{\parbox{2in}{$-\frac{1}{96}\frac{m^4K}{\mu^8}A$}}\\
\hline \raisebox{1cm}{C3} \epsfig{file=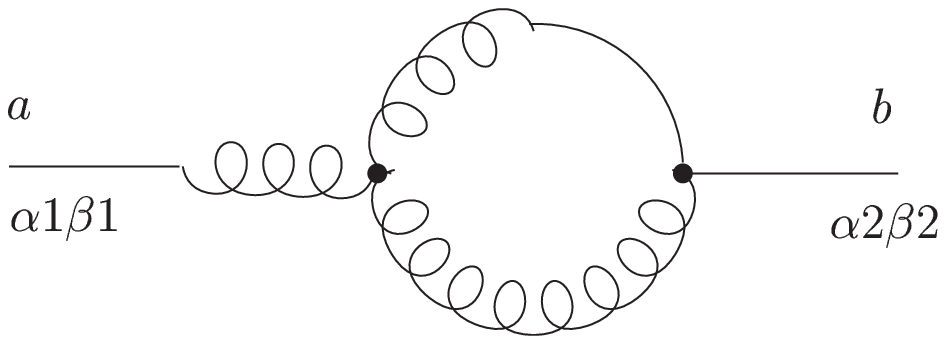, width =2in} &
\raisebox{1cm}{\parbox{.5in}{$1$}} &\raisebox{.5in}{\parbox{2in}{$0$}}\\
\hline \raisebox{1cm}{C4} \epsfig{file=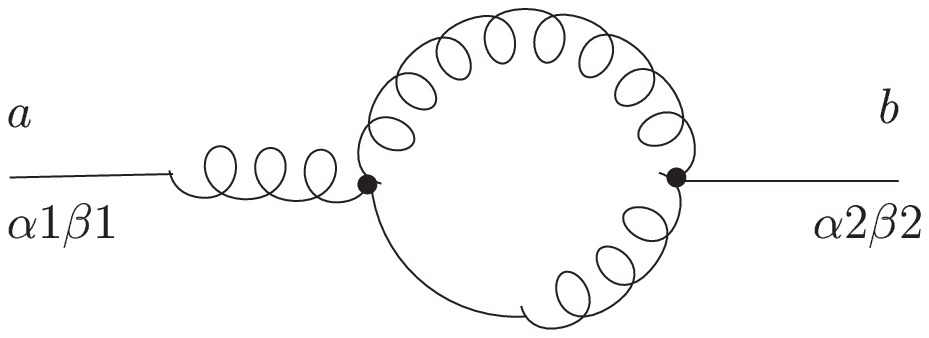, width =2in} &
\raisebox{1cm}{\parbox{.5in}{$1$}}
&\raisebox{.5in}{\parbox{2in}{$-\frac{1}{16}\frac{m^4K}{p^2\mu^8}((p^2+m^2)A-\mu^2B_1)$}}\\
\hline
\end{tabular}
\end{centering}
\end{figure}
\clearpage
\begin{figure}
 \begin{centering}
\begin{tabular}{|c |c |c| }
\hline
    DIAGRAM & SYMMETRY FACTOR & POLE AT $\epsilon=0$\\
    \hline

\hline \raisebox{1cm}{C5} \epsfig{file=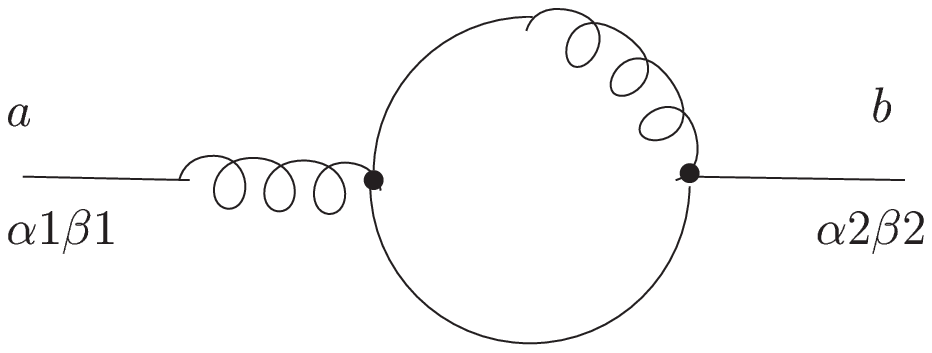, width =2in} &
\raisebox{1cm}{\parbox{.5in}{$1$}} &
\raisebox{.5in}{\parbox{2in}{$0$}}\\
\hline \raisebox{1cm}{C6} \epsfig{file=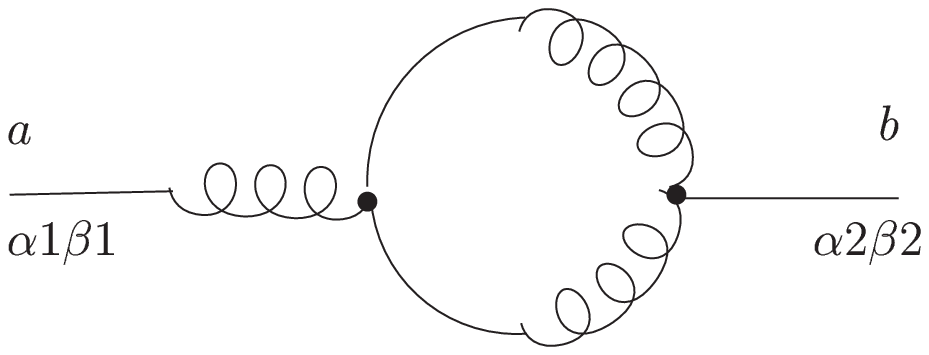, width =2in} &
\raisebox{1cm}{\parbox{.5in}{$1/2$}} &\raisebox{.5in}{\parbox{2in}{$0$}}\\

     \hline \raisebox{1cm}{C7} \epsfig{file=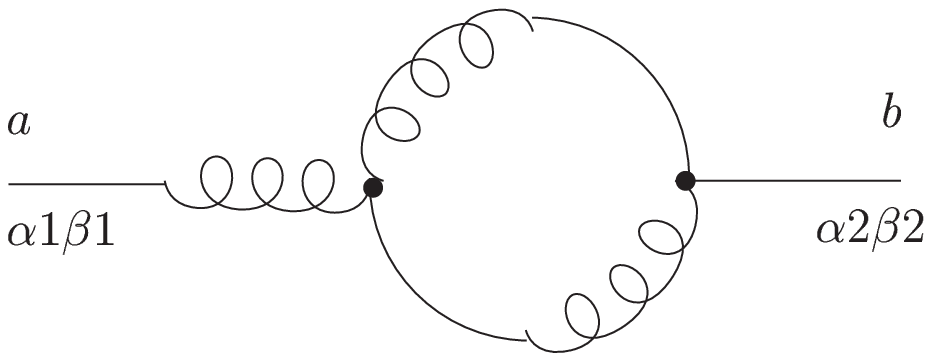, width =2in} &
     \raisebox{1cm}{\parbox{.5in}{$1$}} &\raisebox{.5in}{\parbox{2in}{$-\frac {5}{96}\frac{m^4K}{\mu^8}A$}}\\
     \hline \raisebox{1cm}{D1} \epsfig{file=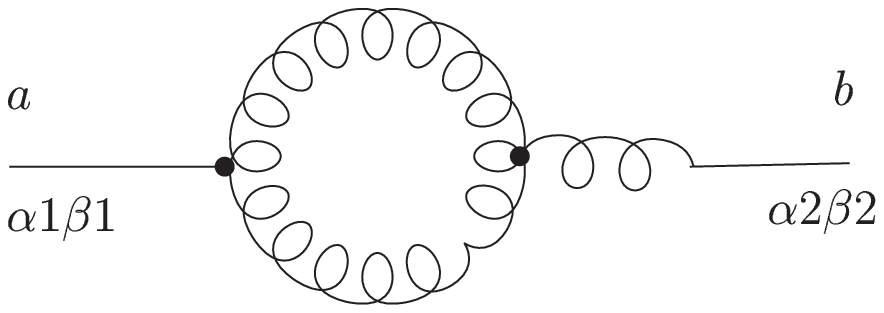, width =2in} &
\raisebox{1cm}{\parbox{.5in}{$1/2$}} &\raisebox{.5in}{\parbox{2in}{$\frac {3}{32}\frac{m^2K}{p^2\mu^4}A$}}\\
\hline \raisebox{1cm}{D2} \epsfig{file=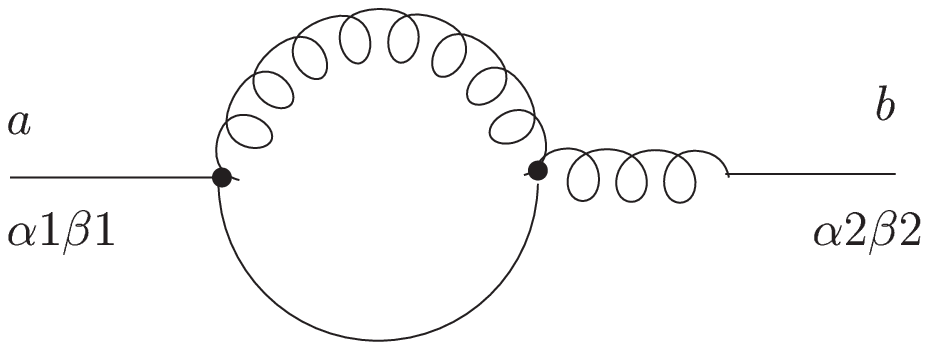, width =2in} &
\raisebox{1cm}{\parbox{.5in}{$1$}} &\raisebox{.5in}{\parbox{2in}{$-\frac {1}{96}\frac{m^4K}{\mu^8}A$}}\\
\hline \raisebox{1cm}{D3} \epsfig{file=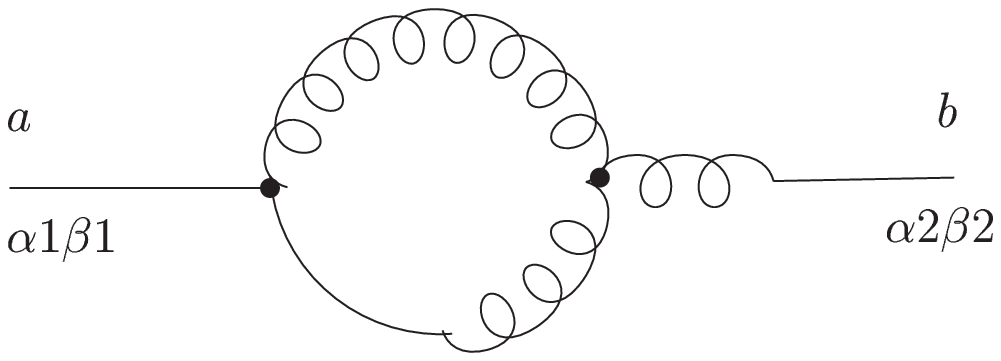, width =2in} &
\raisebox{1cm}{\parbox{.5in}{$1$}}
&\raisebox{.5in}{\parbox{2in}{$0$}}\\
\hline \raisebox{1cm}{D4} \epsfig{file=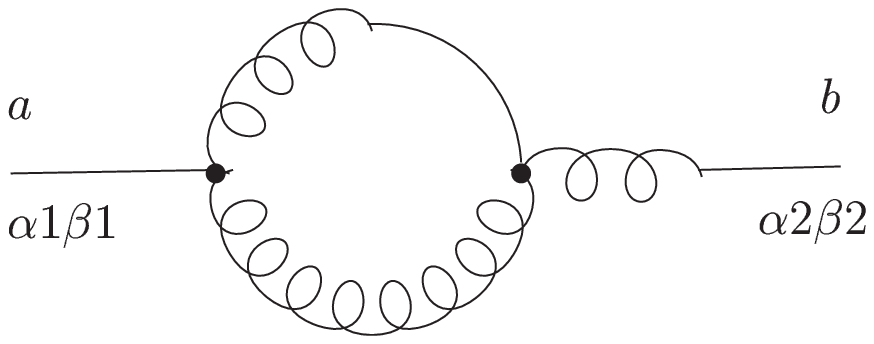, width =2in} &
\raisebox{1cm}{\parbox{.5in}{$1$}}
&\raisebox{.5in}{\parbox{2in}{$-\frac{1}{16}\frac{m^4K}{p^2\mu^8}((p^2+m^2)A-\mu^2B_2)$}}\\
\hline \raisebox{1cm}{D5} \epsfig{file=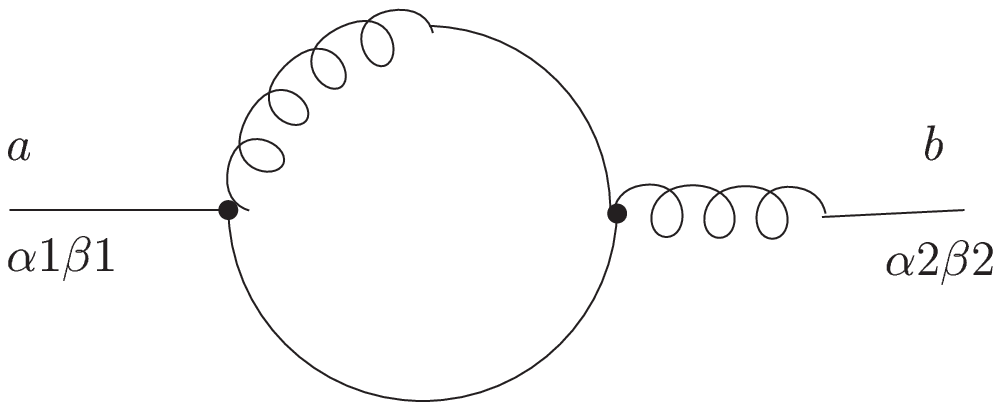, width =2in} &
\raisebox{1cm}{\parbox{.5in}{$1$}} &\raisebox{.5in}{\parbox{2in}{$0$}}\\
\hline \raisebox{1cm}{D6} \epsfig{file=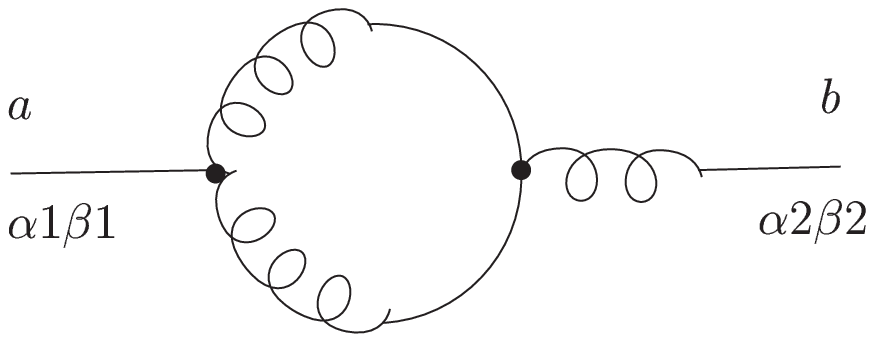, width =2in} &
\raisebox{1cm}{\parbox{.5in}{$1/2$}} &\raisebox{.5in}{\parbox{2in}{$0$}}\\
\hline \raisebox{1cm}{D7} \epsfig{file=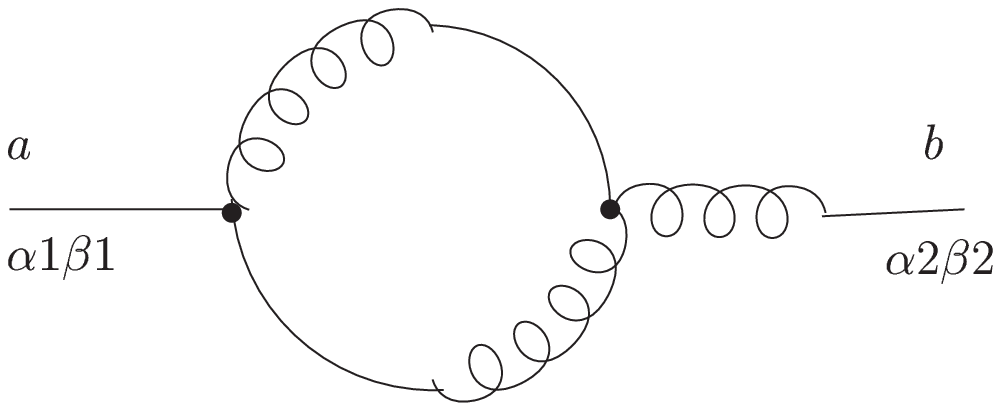, width =2in} &
\raisebox{1cm}{\parbox{.5in}{$1$}}
&\raisebox{.5in}{\parbox{2in}{$\frac{5}{96}\frac{m^4K}{\mu^8}A$}}\\
\hline
\end{tabular}
\caption{One loop contributions to
$<{\phi^{a}}_{\alpha1\beta1}(p){\phi^{b}}_{\alpha2\beta2}(-p)>$.}
\end{centering}
\end{figure}
\clearpage The diagrams give, in total, a pole contribution to the
one loop tensor self energy equal to
\begin{equation}
\frac{m^2K}{48p^2\mu^8}[(12p^4+5p^2m^2+14\mu^4)A-6\mu^2p^2(B_1+B_2)+3p^2\mu^4C].
\end{equation}
The form of this divergence is distinct from the form of the
kinetic term for the tensor field in eq.(1). We note that the
shift ${\phi^{a}}_{\mu \nu}={\chi^{a}}_{\mu
\nu}-\frac{m}{\mu^2}{F^{a}}_{\mu \nu}$ allows us to rewrite eq.
(1) as
\begin{equation}
L=-\frac{1}{4}F_{\mu\nu}^aF_{\mu\nu}^a+\frac{1}{12}H_{\mu\nu\lambda}^a
H_{\mu\nu\lambda}^a +
\frac{\mu^2}{8}\epsilon_{\mu\nu\lambda\sigma}\chi_{\mu\nu}^a
\chi_{\lambda\sigma}^a+\frac{m^2}{8\mu^2}\epsilon_{\mu\nu\lambda\sigma}F_{\mu\nu}^a
F_{\lambda\sigma}^a
\end{equation}
where
\begin{equation}
H_{\mu\nu\lambda}^a=D_{\mu}^{ab}\chi_{\nu\lambda}^b
+D_{\nu}^{ab}\chi_{\lambda\mu}^b +D_{\lambda}^{ab}\chi_{\mu\nu}^b
\end{equation}
(we have employed the Bianchi identity
$\epsilon_{\mu\nu\lambda\sigma}D_{\nu}^{ab}F_{\lambda\sigma}^b=0$
in arriving at eq.(5)). The Feynman rules associated with $L$ in
eq.(5) are identical to the $m \rightarrow 0$ limit of the Feynman
rules associated with the lagrangian $L$ in eq.(1). As a result,
we see that as eq.(4) vanishes when $m = 0$, the two point
function
$<{\chi^{a}}_{\alpha1\beta1}(p){\chi^{b}}_{\alpha2\beta2}(-p)>$
should vanish. Having computed
$<{\phi^{a}}_{\alpha1\beta1}(p){\phi^{b}}_{\alpha2\beta2}(-p)>$
and $<{F^{a}}_{\alpha1\beta1}(p){F^{b}}_{\alpha2\beta2}(-p)>$ we
can thus infer that
$<{\phi^{a}}_{\alpha1\beta1}(p){F^{b}}_{\alpha2\beta2}(-p)>$
should cancel the contribution of eq.(4) in the evaluation of
$<{\chi^{a}}_{\alpha1\beta1}(p){\chi^{b}}_{\alpha2\beta2}(-p)>$.
We hope to explicitly verify this. \\
 Computation of the finite parts of the diagrams of Figure
(1) is complicated by having to assign a meaning to
$\epsilon_{\mu\nu\alpha\beta}\epsilon_{\mu\nu\gamma\delta}$
outside of four dimensions. The $n$ dimensional result for this
differs from the four dimensional result by terms of order
$(n-4)$; consequently when this expression arises in the course of
computing one of the divergent diagrams in Figure (1), the finite
part receives an ill defined contribution. The ambiguity in
$\epsilon_{\mu\nu\alpha\beta}\epsilon_{\mu\nu\gamma\delta}$
cannot, however, affect the pole piece of these diagrams. This
problem does not arise in ref.[2] as there all diagrams
contributing to the vector self energy that involve
$\epsilon_{\mu\nu\lambda\sigma}$ sum  to zero prior to
evaluating any loop momentum integrals.\\
 \\
\textbf{Acknowledgements}\\
NSERC provided financial help. AB's research at Perimeter
Institute is supported in part by the Government of Canada through
NSERC and by the Province of Ontario through MEDT. Roger Macloud
had a helpful suggestion.


\vspace{9pt}

\end{document}